\documentclass[twocolumn,floatfix,prl,showpacs,superscriptaddress]{revtex4}
\usepackage{graphicx}
\usepackage{times,xspace}
\usepackage{amsbsy,amssymb,amsmath,bm}
\usepackage{graphicx,color,epsfig,rotate}
\usepackage{fancyhdr}

\def\bbbc{{\mathchoice {\setbox0=\hbox{$\displaystyle\rm C$}\hbox{\hbox
to0pt{\kern0.4\wd0\vrule height0.9\ht0\hss}\box0}}
{\setbox0=\hbox{$\textstyle\rm C$}\hbox{\hbox
to0pt{\kern0.4\wd0\vrule height0.9\ht0\hss}\box0}}
{\setbox0=\hbox{$\scriptstyle\rm C$}\hbox{\hbox
to0pt{\kern0.4\wd0\vrule height0.9\ht0\hss}\box0}}
{\setbox0=\hbox{$\scriptscriptstyle\rm C$}\hbox{\hbox
to0pt{\kern0.4\wd0\vrule height0.9\ht0\hss}\box0}}}}

\begin{document}
\title{Magnetic Excitations in the Spin-1 Anisotropic Heisenberg Antiferromagnetic Chain System NiCl$_2$-4SC(NH$_2$)$_2$}
\author{S. A. Zvyagin}
\affiliation{Dresden High Magnetic Field Laboratory (HLD),
Forschungszentrum Dresden - Rossendorf, 01314 Dresden, Germany}
\author{J. Wosnitza}
\affiliation{Dresden High Magnetic Field Laboratory (HLD),
Forschungszentrum Dresden - Rossendorf, 01314 Dresden, Germany}
\author{C. D. Batista}
\affiliation{Theoretical Division, Los Alamos National Laboratory,
Los Alamos, NM 87545, USA}
\author {M. Tsukamoto}
\affiliation {Institute for Solid State Physics, University of
Tokyo, Kashiwa, Chiba 227-8581, Japan}
\author {N. Kawashima}
\affiliation {Institute for Solid State Physics, University of
Tokyo, Kashiwa, Chiba 227-8581, Japan}
\author{J. Krzystek}
\affiliation{National High Magnetic Field Laboratory, Florida
State University, Tallahassee, FL 32310, USA}
\author{V. S. Zapf}
\affiliation{National High Magnetic Field Laboratory, Los Alamos
National Laboratory, MS-E536, Los Alamos, NM 87545, USA}
\author{M. Jaime}
\affiliation{National High Magnetic Field Laboratory, Los Alamos
National Laboratory, MS-E536, Los Alamos, NM 87545, USA}
\author{N. F. Oliveira, Jr.}
\affiliation{Instituto de Fisica, Universidade de Sao Paulo,
05315-970 Sao Paulo, Brazil}
\author{A. Paduan-Filho}
\affiliation{Instituto de Fisica, Universidade de Sao Paulo,
05315-970 Sao Paulo, Brazil}

\begin{abstract}  NiCl$_2$-4SC(NH$_2$)$_2$ (DTN) is a  quantum
$S=1$ chain system with strong easy-pane anisotropy and a new
candidate for the Bose-Einstein condensation of the spin degrees
of freedom. ESR studies of magnetic excitations in DTN in fields
up to 25 T are presented. Based on analysis of the  single-magnon
excitation  mode in the high-field spin-polarized phase and
previous experimental results [Phys. Rev. Lett. 96, 077204
(2006)], a revised set of spin-Hamiltonian parameters  is
obtained. Our results yield $D=8.9$ K, $J_c=2.2$ K, and
$J_{a,b}=0.18$ K for the anisotropy, intrachain, and interchain
exchange interactions, respectively. These values are used to
calculate the antiferromagnetic phase boundary, magnetization and
the frequency-field dependence of two-magnon bound-state
excitations predicted by theory and observed in DTN for the first
time. Excellent quantitative agreement with experimental data is
obtained.

\end{abstract}

\maketitle

 Antiferromagnetic (AFM) quantum spin-1 chains have been the
subject of intensive theoretical and experimental studies,
fostered especially by the Haldane conjecture \cite{Haldane1-DTN}.
Due to quantum fluctuations, an isotropic spin-1 chain has a
spin-singlet ground state separated from the first excited state
by a gap $\Delta \sim 0.41 J$ \cite{Botet-DTN}, where $J$ is the
exchange interaction. As shown by Golinelli $et~al.$
\cite{Golinelli-DTN}, the presence of a strong easy-plane
 anisotropy $D$ can significantly modify the excitation
spectrum, so that the gap size is not determined by the strength
of the AFM quantum fluctuations $exclusively$, but depends on the
dimensionless parameter $\rho=D/J$. The Haldane phase is predicted
to survive up to $\rho_c = 0.93$ \cite{Sakai-DTN}, where the
system undergoes a quantum phase transition.  For $\rho
> \rho_c$ the gap reopens, but its origin is dominated by the
anisotropy $D$, and the system is in the so-called large-$D$
regime.  While the underlying physics of Haldane chains is fairly
well understood, relatively little is known about the magnetic
properties (and particularly the elementary excitation spectrum)
of non-Haldane $S=1$ AFM chains in the large-$D$ phase. Intense
theoretical work and numerous predictions
\cite{Golinelli-DTN,Sakai-DTN, Silberglitt-DTN, Chiu-DTN,
Papanicolaou1-DTN, Orendac1-DTN, Papanicolaou-DTN, Kolezhuk-DTN}
make the experimental investigation of large-$D$ spin-1 chains a
topical problem in low-dimensional magnetism.

Recently, weakly-coupled spin-1 chains have attracted renewed
interest due to their possible relevance to the field-induced
Bose-Einstein condensation (BEC) of magnons. When the field $H$,
applied perpendicular to the easy plane, exceeds a critical value
$H_{c1}$ (defined at $T=0$), the gap closes and the system
undergoes a transition into an $XY$-like AFM phase with a finite
magnetization and AFM magnon excitations. If the spin Hamiltonian
has axial symmetry with respect to the applied field, the AFM
ordering can be described as BEC of magnons by mapping the spin-1
system into a gas of semi-hard-core bosons \cite{Batista2-DTN}.
The applied field plays the role of a chemical potential, changing
the boson population.  In accordance with mean-field BEC theory
\cite{Affleck-DTN, Giamarchi-DTN, Nikuni-DTN}, the phase-diagram
boundary for a three-dimensional system should obey a power-law
dependence, $H - H_{c1} \sim T^{3/2}_c$, as $T\rightarrow 0$.
Above the upper critical field, $H_{c2}$ (defined at $T=0$), the
system is in a fully spin-polarized (FSP) phase,  and the
excitation spectrum is formed by gapped magnons.

The compound NiCl$_2$-4SC(NH$_2$)$_2$ (dichloro-$tetrakis$
thiourea-nickel(II), known as DTN), is a new candidate for
studying the BEC phenomenon in magnetic fields. It has a
tetragonal crystal structure, space group $I$4, with two molecules
in the unit cell \cite{Paduan2-DTN, Paduan-DTN}. The anisotropy,
intra- and inter-chain exchange parameters, $D=8.12$ K, $J_c=1.74$
K, and $J_{a,b}=0.17$ K, respectively, were obtained from  a fit
of zero-field inelastic-neutron-scattering data \cite{Zapf-DTN}
using generalized spin-wave theory \cite{Copper}. It was shown
that the Ni spins are strongly coupled along the tetragonal axis,
making DTN a system of weakly interacting $S=1$ chains with
single-ion anisotropy larger than the intra-chain exchange
coupling. Recently, it was proposed \cite{Paduan-DTN,Zapf-DTN}
that the field-induced low-temperature transition of DTN to the
AFM-ordered state can be interpreted as a BEC of magnons,
 with values of $H_{c1}=2.1$ T and
$H_{c2}=12.6$ T for the lower and upper critical fields,
respectively. Although the observed overall picture is consistent
with the BEC scenario, some questions remain. One of the unsolved
problems is a pronounced (about 2 T) disagreement  between the
 value for the upper critical field obtained experimentally and
the predicted one ($H_{c2}=10.85$ T) \cite{Zapf-DTN}. This
discrepancy is rather puzzling, particularly in light of the
excellent agreement for $H_{c1}$.

In this Letter, we report  electron spin resonance (ESR) studies
of the elementary excitation spectrum in DTN in magnetic fields up
to 25 T ($\sim 2H_{c2}$). A distinct advantage of the high-field
approach for determining the spin-Hamiltonian parameters of spin-1
AFM chains is the availability of $exact$ theoretical expressions
for the spin-polarized phase. Analysis of the high-field
single-magnon branch has allowed us to obtain the precise value
for the anisotropy parameter, $D=8.9$ K. Using results of recent
thermodynamic and neutron-scattering measurements
\cite{Paduan-DTN,Zapf-DTN}, we were able to refine the exchange
parameters, obtaining $J_c=2.2$ K, $J_{a,b}=0.18$ K for intra- and
inter-chain exchange interactions, respectively. In addition, we
present magnetocaloric-effect and low-temperature magnetization
data allowing us to check the obtained set of parameters. These
values agree well with those obtained by fitting the AFM phase
boundary and the magnetization with results of quantum Monte Carlo
simulations, and nicely reproduce both critical fields,
$H_{c1}=2.1$ T and $H_{c2}=12.6$ T. Furthermore, we report on the
first direct and reliable observation of the two-magnon bound
states in spin-1 AFM chain system with strong easy-plane
anisotropy, predicted by theory \cite{Papanicolaou-DTN} for the
high-field spin-polarized phase. Using the obtained parameters of
the spin-Hamiltonian we were able to calculate frequency-field
dependences of the two-magnon bound state excitations. Excellent
quantitative agreement between the theoretical predictions and
experiment is obtained.

The investigation of the  magnetic excitation spectra in
single-crystalline DTN samples  was done using a tunable-frequency
submillimeter-wave ESR spectrometer \cite{Zvyagin-DTN} with
external field $H$  applied along the tetragonal $c$ axis.

\begin{figure}
\begin{center}
\vspace{2.6cm}
\includegraphics[width=0.45\textwidth]{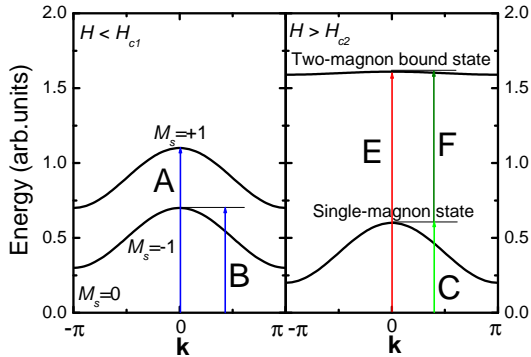}
\vspace{-2.8 cm}
\caption{\label{fig:Disp-DTN}  (color online) A schematic view of
the magnetic-excitation dispersion in an $S=1$ Heisenberg chain
with strong easy-plane ($D>0$) anisotropy for two arbitrary
fields, $H<H_{c1}$ (left) and $H>H_{c2}$ (right), with a magnetic
field applied along the principal axis $z$. Note that the ESR
transitions denoted by A, B, C, E, and F occur at \textbf{k} = 0.
Two-particle continua predicted for both regions are not shown for
simplicity.}
\end{center}
\end{figure}

\begin{figure}
[!b]
\begin{center}
\vspace{3.2cm}
\includegraphics [width=0.55\textwidth] {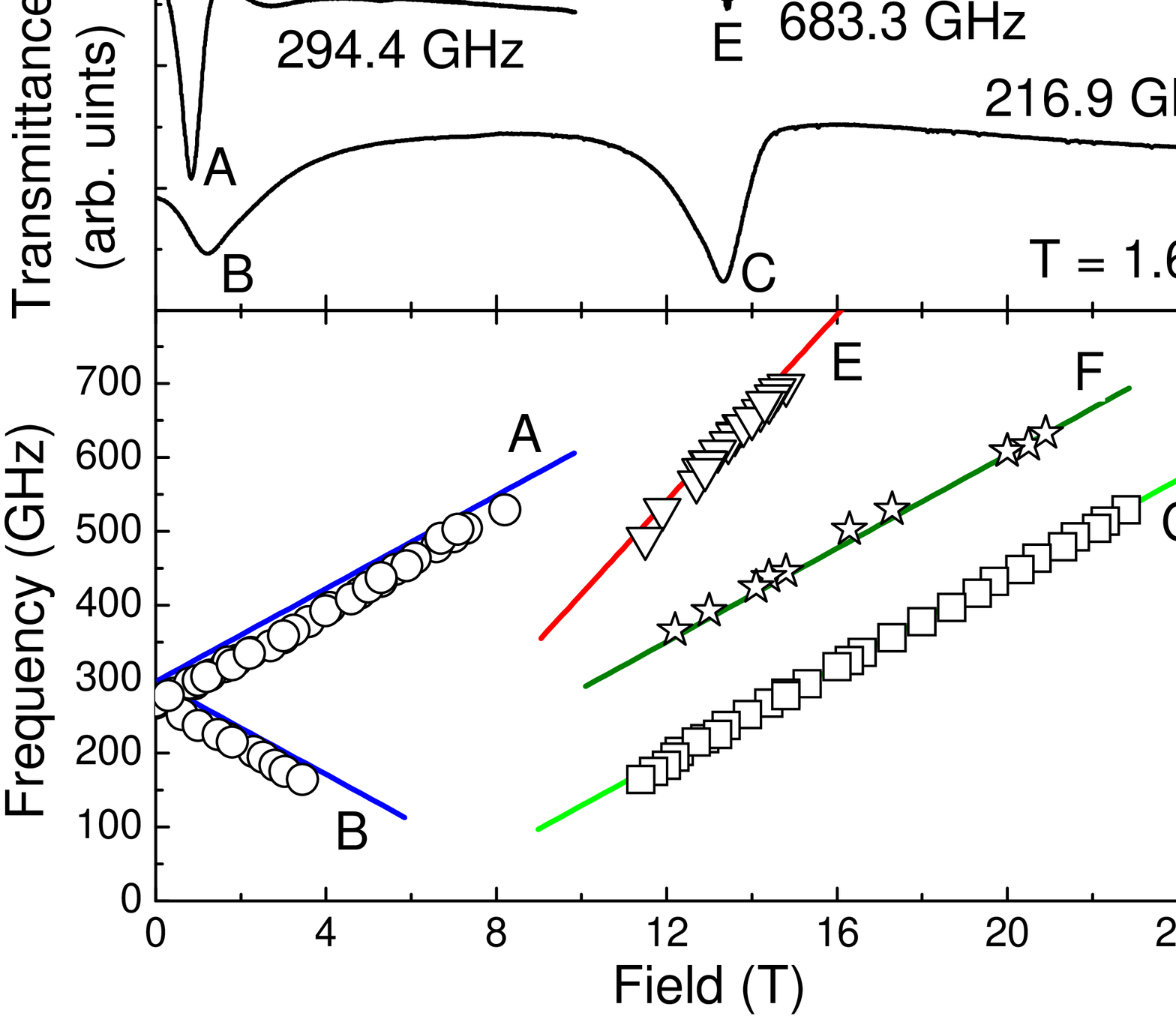}
\vspace{-3.4 cm} \caption{\label{fig:DTN-FFD-SP} (color online)
Top: Typical ESR transmission spectra in DTN, taken at frequencies
of 216.9, 294.4, and 683.3 GHz at $T=1.6$ K, $H~\|~c$. Bottom:
Frequency-field dependence of magnetic excitations in DTN, taken
at temperatures of 1.6 K (the modes A, B, C, and E) and 4.3 K (the
mode F), $H~\|~c$. Symbols denote the experimental results, and
lines correspond to results of calculations (see text for
details).}
\end{center}
\end{figure}

In sufficiently small fields, a spin-1 Ni(II) system  is in the
quantum-paramagnetic (QPM) phase, having the $M_s=0$ ground state.
The excitation spectrum  is gapped and determined by the $\Delta
M_s = \pm1$ transitions. For fields $H$ applied perpendicular to
the easy plane, the Hamiltonian can be written as
\begin{eqnarray}
\label{Ham}
\mathcal{H}&=&\sum_{\textbf{j},\nu}
J_{\nu}\textbf{S}_{\textbf{j}}\cdot
\textbf{S}_{\textbf{j}+e_{\nu}}+ \sum_{\textbf{j}} [D (S^z_\textbf{j})^2+g\mu_B
H  S^z_\textbf{j}],
\end{eqnarray}
where $\nu=\{a,b,c\}$. A corresponding dispersion of low-field
magnetic excitations is shown schematically in Fig.\
\ref{fig:Disp-DTN} (left). Then, the frequency-field dependence of
the  $\Delta M_s = \pm1$ ESR transitions can be written as
$\omega^{A/B}=\Delta\pm g\mu_B H$, where $\Delta$ corresponds to
the energy gap at \textbf{k} = 0 and $H=0$. These transitions were
observed in our experiments and are denoted in Fig.\
\ref{fig:DTN-FFD-SP} by circles \cite{impurity}. The zero-field
splitting measured directly yields $\Delta=269$ GHz, which agrees
quite well with results of neutron-scattering measurements,
$\Delta_{\textbf{k}=0}=1.1$ meV \cite{Zapf-DTN}. No splitting
within the excited doublet in zero field was observed, which
indicates the absence of  in-plane anisotropy. From the slope of
the frequency-field dependences of the branches A and B, the $g$
factor was determined directly, $g=2.26$ \cite{accuracy}.

As mentioned before,  when the magnetic field exceeds the upper
critical field, $H_{c2}$, the system is in the fully
spin-polarized state (a corresponding dispersion of low-energy
magnetic excitations is shown  schematically in Fig.\
\ref{fig:Disp-DTN} (right)). Excitations from the ground state to
the single-magnon state have been observed in our experiments and
are denoted by squares in Fig.\ \ref{fig:DTN-FFD-SP} (the
resonance C). These excitations correspond to a single-spin flip
from the $S_z=-1$ to $S_z=0$ state and are uniformly delocalized
over the entire lattice with a well-defined momentum \textbf{k}.
The $S_z=0$ state propagates along the lattice as a free
quasiparticle with hopping $J_{\nu}$ along the $\nu$ direction
($\nu=\{a,b,c\}$), that arises from the transverse part of the
Heisenberg interaction. There are also diagonal energy gains of
$-2(J_c + 2J_a)$  due to the Ising part, and $-D$ due to the
single-ion anisotropy. The diagonal energy cost comes from the
Zeeman interaction $g\mu_BH$. Then, the single-magnon excitation
dispersion can be calculated $exactly$:
\begin{eqnarray}
\label{Disp-magnons}\omega(\textbf{k}) = g\mu_BH - D - 2(J_c+2J_a)
+ 2(J_c \text{cos}k_z + \nonumber\\J_a \text{cos}k_x  + J_b
\text{cos}k_y).
\end{eqnarray}
The ESR transitions taking place at $\textbf{k}=0$ have the
frequency $\label{FFD-high-C}\omega_{C}=g\mu_B H-D$. The best fit
of the ESR data denoted by squares in Fig.\ \ref{fig:DTN-FFD-SP}
reveals $D=8.9$ K for the anisotropy constant. From the exact
expression for $H_{c2}$, given as \cite{Zapf-DTN}
\begin{eqnarray}
\label{Hc2} H_{c2}=\frac{1}{g\mu_B}(D + 4\sum_{\eta}J_{\eta}),
\end{eqnarray}
and using $H_{c2}=12.6$ T \cite{Paduan-DTN, Zapf-DTN} we obtain
$\sum_{\eta} J_{\eta}= J_c + 2J_a=2.557$ K. The zero-field
dispersion of magnetic excitations calculated using
neutron-scattering data \cite{Zapf-DTN} yields $J_a/J_c=0.082$.
Thus, in addition to the anisotropy constant $D=8.9$ K, all three
exchange parameters, $J_c=2.2$ K and $J_a=J_b=0.18$ K, can be
calculated quite precisely.

\begin{figure}
\begin{center}
\vspace{2.6cm}
\includegraphics[width=0.5\textwidth]{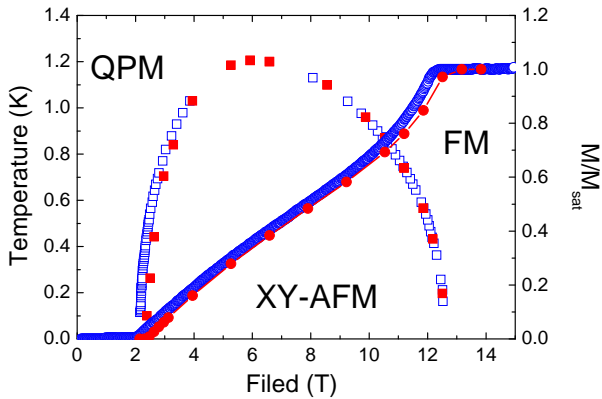}
\vspace{-3.2 cm} \caption{\label{fig:DTN-FD-M} (color online)
Temperature-field diagram of the AFM-ordered phase obtained from
magnetocaloric-effect measurements (opened squares), and the
magnetization data taken at $T=16$ mK (opened circles), $H~\|~c$.
Results of the quantum Monte Carlo calculations of the
phase-diagram boundaries and the magnetization using the set of
parameters obtained as described in text are denoted by closed
squares and circles, respectively.}
\end{center}
\end{figure}

The phase boundary (obtained from magnetocaloric-effect
measurements) and the field dependence of the magnetization at
$T=16$ mK were computed for the obtained set of parameters by
means of a quantum Monte Carlo simulations for a finite lattice of
$L^3$ sites, $L=16$. Fig.\ \ref{fig:DTN-FD-M}  shows a very good
agreement between the calculated (closed symbols) and experimental
(opened symbols) data.

In addition to ordinary single-magnon states and two-magnon
continuum,  the theory \cite{Papanicolaou-DTN} predicts the
existence of two-magnon bound states (sometimes referred to as
single-ion bound states \cite{Silberglitt-DTN}). The physical
picture of the two-magnon bound-state excitations corresponds to a
double-spin-flip transition from $S_z=-1$ to $S_z=+1$. The
transverse part of the Heisenberg term of ${\cal H}$
 mixes this state with the one that has a pair of
$S_z=0$ states. Since the diagonal energy difference between these
two states, $2D$, is much bigger than  $J_c$ and $J_a$ (associated
with hopping in the $c$ and $a$ directions), the distance between
the two $S_z=0$ sites remains finite, giving rise to a two-magnon
bound state.  The two-magnon bound states appear to be a specific
feature of anisotropic spin-1 Heisenberg systems. It is worth
mentioning that the two-magnon bound states were already predicted
in 1970 by Silberglitt and Torrance \cite{Silberglitt-DTN} for
Heisenberg ferromagnets with single-ion anisotropy. Later on, this
subject attracted a great deal of attention due to its potential
relevance  to the intrinsic localized spin modes in anisotropic
ferromagnets \cite{Wallis-DTN} and antiferromagnets
\cite{Lai-DTN}.  It was suggested \cite{Papanicolaou-DTN} that the
two-magnon bound states should make a distinct contribution to the
excitation spectrum of S = 1 large-$D$ AFM chains above the upper
critical field $H_{c2}$ and that their effect can be unambiguously
identified by ESR measurements. A signature of two-magnon bound
states was obtained by means of high-field ESR in the spin-1 chain
compound Ni(C$_2$H$_8$N$_2$)$_2$Ni(CN)$_4$ (known as NENC)
\cite{Zvyagin1-DTN}. A broad absorption was detected in the
high-field spin-polarized phase. Based on analysis of the
temperature dependence of the ESR intensity, this feature was
interpreted as transitions from the single-magnon to two-magnon
bound states. However,  a closer look at ESR spectra in NENC
uncovered a more complicated structure than predicted (including,
for instance, two non-equivalent Ni sites and, most likely, a
finite in-plane anisotropy), which made the accurate quantitative
comparison with the theory difficult.

Here we report on the first observation of transitions from the
ground state to two-magnon bound state in a spin-1 AFM chain
system with strong easy-plane anisotropy in the high-field FSP
phase. The corresponding excitations are denoted by triangles in
Fig.\ \ref{fig:DTN-FFD-SP}, bottom (the resonance E, Fig.\
\ref{fig:DTN-FFD-SP} (top)).  The frequency-field dependence of
the ground-state two-magnon bound-state excitations can be
calculated exactly using the set of parameters obtained as
described above. Results of corresponding calculations are shown
in Fig.\ \ref{fig:DTN-FFD-SP} (bottom) by the line E. One more
resonance absorption  was observed at higher temperatures (which
indicates transitions within excited states). The corresponding
data obtained at $T=4.3$ K are denoted in Fig.\
\ref{fig:DTN-FFD-SP} (bottom) by stars. This ESR mode corresponds
to transitions from the single-magnon to two-magnon bound states
(Fig.\ \ref{fig:Disp-DTN} (right)), which occur at $\textbf{k}=0$.
The frequency-field dependence of these transitions can be
calculated, using the expression $\omega_F=\omega_E - \omega_C $
(where $\omega_E$ and $\omega_C$ are the excitation frequencies
for the modes E and C, respectively), and is denoted in Fig.\
\ref{fig:DTN-FFD-SP} by the line F.  In both cases, excellent
agreement with experimental data was achieved, which shows that
our model actually works fairly well  up to $T\sim J$.

Let us now discuss the low-field quantum-paramagnetic phase. As
mentioned before, in this phase the system has  the $M_s=0$ ground
state and a gapped excitation spectrum  formed by $\Delta M_s =
\pm1$ transitions. The obtained set of spin-Hamiltonian parameters
can be used to calculate the lower critical field, employing the
spin-wave theoretical approach \cite{Copper}:
\begin{eqnarray}
\label{Hc1} H_{c1}=\frac{1}{g\mu_B}\sqrt{\mu^2 - 4 s^2 \mu
\sum_{\eta}J_{\eta}}.
\end{eqnarray}
The parameters $\mu=10.3$ and $s^2=0.92$ can be obtained from the
self-consistent equations:
\begin{eqnarray}
\label{mu-D}D = \mu
(1+\frac{1}{N_s}\sum_{\textbf{k}}\frac{\gamma_\textbf{k}}{\omega_\textbf{k}})
\end{eqnarray}
 and
\begin{eqnarray}
\label{s}s^2 = 2 - \frac{1}{N_s}\sum_{\textbf{k}}\frac{\mu + s^2
\gamma_\textbf{k}}{\omega_\textbf{k}},
\end{eqnarray}
 where   $\gamma_\textbf{k }= 2 \sum_{\nu}J_{\nu}\cos
k_{\nu}$ and $\omega_\textbf{k} = \omega^{A/B}_\textbf{k} (H =
0)$.  It is important to mention, that the disagreement between
the calculated ($H_{c1}=2$ T) and experimental ($H_{c1}=2.1$ T)
values is due to limitations of the spin-wave theory applied to
quasi-one-dimensional spin systems. Using the spin-wave theory
\cite{Copper} and the set of parameters obtained as described
above, the frequency-field dependence of the ESR transitions
 can be calculated using the expression \cite{Zapf-DTN}
\begin{eqnarray}
\label{ffd-low-field} \omega^{A/B} = \sqrt{\mu^2 + 4 s^2 \mu
\sum_{\eta} J_{\eta}} \pm g\mu_B H.
\end{eqnarray}
 Eq.\ (\ref{ffd-low-field}) predicts a value of $H_{c1}=2$ T,
that is in fact in relatively good agreement with the experimental
value $H_{c1}=2.1$ T. The results of the calculations are
presented by the lines A and B in Fig.\ \ref{fig:DTN-FFD-SP}
(bottom) together with experimental data denoted by circles.  The
agreement in this case is reasonably good, considering that the
quasi-one-dimensional nature of the system causes pronounced
quantum fluctuations of the AFM-order parameter in the low-field
QPM phase.

In summary, a systematic ESR study of the magnetic excitations in
DTN, an $S=1$ Heisenberg AFM chain material in the large-D regime,
has been presented. Investigation of the high-field magnon
excitations allowed us to obtain a reliable set of the
spin-Hamiltonian parameters, which was employed to calculate the
ESR spectrum in a broad range of magnetic fields and frequencies.
These values agree very well with the ones obtained from fitting
the AFM-phase boundary and low-temperature magnetization of DTN
with results of quantum Monte Carlo simulations, including both
critical fields. The parameters were used to calculate the
frequency-field dependence of two-magnon bound-state excitations,
predicted by theory and observed in DTN for the first time.
Excellent agreement between the theory and experiment was
obtained.

\emph{Acknowledgments.--} The authors express their sincere thanks
to  A.K. Kolezhuk,  A. Orend\'{a}\v{c}ov\'{a}, and S. Hill for
fruitful discussions. A portion of this work was performed at the
National High Magnetic Field Laboratory, Tallahassee, FL, which is
supported by NSF Cooperative Agreement No. DMR-0084173, by the
State of Florida, and by the DOE. S.A.Z. acknowledges the support
from the NHMFL through the VSP No. 1382. The Monte Carlo
simulation was carried out on SGI Altrix 3700Bx2 at the
Supercomputer Center, ISSP, University of Tokyo. A.P.F and N.F.O.,
Jr. are grateful for support from CNPq and FAPESP (Brazil).


\begin{thebibliography}{10}
\bibitem{impurity} A minor impurity phase was detected in some ESR
spectra. Although the origin of this phase is not clear at the
moment, we speculate that it is most likely originated from a
superficial layer of DTN-crystals attacked by the GE-varnish
solvent, which was used to fix the sample inside the
sample-holder.
\bibitem{accuracy} With accuracy better than $\pm 0.5\%$.  For further references:
the anisotropy constant $D$ is calculated with accuracy better
than $\pm 3\%$; the accuracy of estimation of exchange parameters
is basically determined by accuracy of estimating the upper
critical field $H_{c2}$ and by accuracy of neutron scattering
measurements \cite{Zapf-DTN}, which is better than $\pm 10\%$.
\end{thebibliography}
\end{document}